\documentclass{article}
\usepackage{arxiv}
\usepackage[utf8]{inputenc} % allow utf-8 input
\usepackage[T1]{fontenc}    % use 8-bit T1 fonts
\usepackage{hyperref}       % hyperlinks
\usepackage[lighttt]{lmodern}
\hypersetup{
    colorlinks=true,
    linkcolor=cyan,
    filecolor=magenta,      
    urlcolor=blue,
}
\usepackage{url}            % simple URL typesetting
\usepackage{booktabs}       % professional-quality tables
\usepackage{amsfonts}       % blackboard math symbols
\usepackage{nicefrac}       % compact symbols for 1/2, etc.
\usepackage{microtype}      % microtypography
\usepackage{lipsum}
\usepackage{graphicx}
\usepackage{amsmath}
\usepackage{multirow}
\usepackage{xspace}

\author{
  Yutaro Iiyama \\
  University of Tokyo \\
  Tokyo 113-0033, Japan
  \And
  Gianluca Cerminara, Abhijay Gupta, Jan Kieseler, Vladimir Loncar\thanks{Also at Institute of Physics Belgrade, Pregrevica 118, Belgrade, Serbia}, \\ \textbf{Maurizio Pierini, Shah Rukh Qasim\thanks{Also at Manchester Metropolitan University, Manchester M15 6BH, UK}, Marcel Rieger, Sioni Summers,}\\
  \textbf{Gerrit Van Onsem, Kinga Anna Wozniak\thanks{Also at University of Vienna, 1010 Vienna, Austria}}\\
  European Organization for Nuclear Research (CERN) \\
  CH-1211 Geneva 23, Switzerland
  \And
  Jennifer Ngadiuba\\
  California Institute of Technology\\
  Pasadena, CA 91125, USA
  \And
  Giuseppe Di Guglielmo \\
  Columbia University \\
  New York, NY 10027, USA
  \And
  Javier Duarte \\
  University of California San Diego \\
  La Jolla, CA 92093, USA \\
  \And 
  Philip Harris, Dylan Rankin\\
  Massachusetts Institute of Technology\\
  Cambridge, MA 02139, USA 
  \And
  Sergo Jindariani, Mia Liu, Kevin Pedro, Nhan Tran\thanks{Also at Northwestern University, Evanston, IL 60208, USA} \\
  Fermi National Accelerator Laboratory\\
  Batavia, IL 60510, USA\\
  \And
  Edward Kreinar\\
  HawkEye360\\
  Herndon, VA 20170, USA\\
  \And
  Zhenbin Wu\\
  University of Illinois at Chicago\\
  Chicago, IL 60607, USA
}

\newcommand{\garnet}{\textsc{Gar\-Net}}
\newcommand{\hlsfml}{\texttt{hls4ml}}
\newcommand{\FIN}{\ensuremath{F_{\mathrm{in}}}}
\newcommand{\FOUT}{\ensuremath{F_{\mathrm{out}}}}
\newcommand{\FLR}{\ensuremath{F_{\mathrm{LR}}}}

\newcommand{\Vmax}{\ensuremath{V_{\mathrm{max}}}}
\newcommand{\Rreuse}{\ensuremath{R_{\mathrm{reuse}}}}
\newcommand{\fprim}{\ensuremath{f_{\mathrm{prim}}}}
\newcommand{\Epred}{\ensuremath{E_{\mathrm{pred}}}}
\newcommand{\Etrue}{\ensuremath{E_{\mathrm{true}}}}
\newcommand{\DEDh}{\ensuremath{\Delta\Epred / \Delta h}}
\newcommand{\millisec}{\ensuremath{\,\mathrm{ms}\xspace}}
\newcommand{\microsec}{\ensuremath{\,\mu\mathrm{s}\xspace}}
\newcommand{\nanosec}{\ensuremath{\,\mathrm{ns}\xspace}}
\newcommand{\keras}{\textsc{Keras}}
\newcommand{\qkeras}{\textsc{QKeras}}
\newcommand{\pytorch}{\textsc{PyTorch}}
\newcommand{\tensorflow}{\textsc{TensorFlow}}
\newcommand{\onnx}{\textsc{ONNX}}

\begin{document}
\title{Distance-Weighted Graph Neural Networks on FPGAs for Real-Time Particle Reconstruction in High Energy Physics}
%preprint
\begin{flushright}
  FERMILAB-PUB-20-405-E-SCD
\end{flushright}
%%%%%%%%%%%%%%%%%%%%%%%%%%%%%%%%%%%
% hls4ml logo
\begin{center}
\includegraphics[width=8cm]{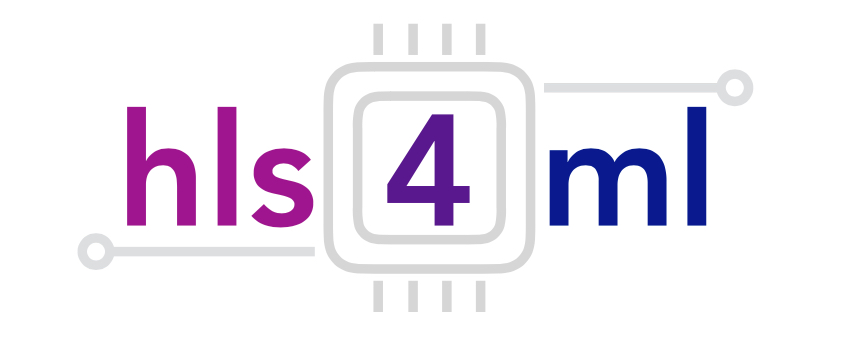}
\end{center}
%%%%%%%%%%%%%%%%%%%%%%%%%%%%%%%%%%%

\maketitle
\newpage

\begin{abstract}
Graph neural networks have been shown to achieve excellent performance for several crucial tasks in particle physics, such as charged particle tracking, jet tagging, and clustering. 
 An important domain for the application of these networks is the  FGPA-based first layer of real-time data filtering at the CERN Large Hadron Collider, which has strict latency and resource constraints. 
 We discuss how to design distance-weighted graph networks that can be executed with a latency of less than {1\microsec} on an FPGA. 
 To do so, we consider a representative task associated to particle reconstruction and identification in a next-generation calorimeter operating at a particle collider. 
 We use a graph network architecture developed for such purposes, and apply additional simplifications to match the computing constraints of Level-1 trigger systems, including weight quantization. 
 Using the {\hlsfml} library, we convert the compressed models into firmware to be implemented on an FPGA. 
 Performance of the synthesized models is presented both in terms of inference accuracy and resource usage.
\end{abstract}

% keywords can be removed

\section{Introduction}

At the CERN Large Hadron Collider (LHC), high-energy physics (HEP) experiments collect signals generated by the particles produced in high-energy proton collisions that occur every {25\nanosec}, when two proton beams cross. 
The readout from the detectors that capture the particles emerging from the collision is filtered by a real-time processing system, known as the \textit{trigger}, that discards uninteresting collision events, based on a set of predefined algorithms. 
The trigger system is structured in two stages: a Level-1 trigger (L1T), implemented with custom electronics on-detector and field-programmable gate arrays (FPGAs); and a high-level trigger (HLT), consisting of a computer farm, possibly including co-processor accelerators like graphics processing units (GPUs) and FPGAs.
Because of asynchronous event processing at the HLT, the accept/reject decision has to be reached with a typical latency of $\mathcal{O}(100)\millisec$. 
However, at the L1T, a decision must be taken within a fixed latency of $\mathcal{O}(1)\microsec$. 
The main limitations are the synchronous, ``hard-deadline'' nature of the processing system and the limited size of the memory buffer for the data from each beam crossing.

While HLT algorithms have a complexity comparable to those used \textit{offline} to produce the final physics results, a typical L1T algorithm consists of simpler rules based on coarser objects to satisfy the latency constraint. 
Consequently, the resolution of quantities computed at the L1T is typically poor compared to offline quantities. 
Recently, the successful deployment of the first machine learning (ML) L1T algorithm, based on a boosted decision tree (BDT), at the LHC~\cite{cms_l1t_bdt} has changed this tendency, raising interest in using ML inference as fast-to-execute approximations of complex algorithms with good accuracy.
This first example consisted of a large, pre-computed table of input and output values implementing a BDT, which raises the question of how to deploy more complex architectures. 
%To this end, currently, one is limited by the compute resources of an FPGA chip, e.g. the number of digital signal processing units (DSPs).
This question motivated the creation of {\hlsfml}~\cite{Duarte:2018ite,hls4ml}, a library designed to facilitate the deployment of ML algorithms on FPGAs. 

A typical {\hlsfml} workflow begins with a neural network model that is implemented and trained using \keras~\cite{chollet2015keras}, \pytorch~\cite{pytorch}, or \tensorflow~\cite{TensorFlow}.
The trained model is passed to {\hlsfml}, directly or through the \onnx~\cite{bai2019} interface, and converted to C++ code that can be processed by a high-level synthesis (HLS) compiler to produce an FPGA firmware. 
By design, {\hlsfml} targets low-latency applications. 
To this end, its design prioritizes all-on-chip implementations of the most common network components. 
Its functionality has been demonstrated with dense neural networks (DNNs)~\cite{Duarte:2018ite}, extended to also support BDTs~\cite{Summers:2020xiy}. 
Extensions to convolutional and recurrent neural networks are in development. 
The library comes with handles to compress the model by quantization, up to binary and ternary precision~\cite{DiGuglielmo:2020eqx}. 
Recently, support for \qkeras~\cite{qkeras} models has been added, in order to allow for quantization-aware training of models~\cite{Coelho:2020zfu}.
While the {\hlsfml} applications go beyond HEP, its development has been driven by the LHC L1T use case.

Graph neural networks (GNNs) are among the complex architectures whose L1T implementations are in high demand, given the growing list of examples showing how well GNNs can deal with tasks related to HEP~\cite{neuraljets,Abdughani:2018wrw,Choma:2018zbe,Martinez:2018fwc,Qasim:2019otl,Qu:2019gqs,Moreno:2019bmu,Moreno:2019neq,Jin:2019cbv,ExaTrkX,Bernreuther:2020vhm,Shlomi:2020gdn}. 
In fact, while the irregular geometry of a typical HEP detector complicates the use of computing vision techniques such as convolutional neural networks, GNNs can naturally deal with the sparse and irregular nature of HEP data.

In this work, we show how a graph model can be efficiently deployed on FPGAs to perform inference within $\mathcal{O}(1)\microsec$ for HEP-related problems. 
We consider the distance-weighted architecture \garnet, introduced in Ref.~\cite{Qasim:2019otl}, which is designed to keep resource consumption under control by reducing as much as possible the number of operations. 
It has been demonstrated to perform well for a HEP-related task, namely particle reconstruction in a calorimeter.
For these reasons, it represents a good candidate for our purpose. 
The firmware implementation of {\garnet} presented in this work has been included in \hlsfml, representing the first graph-based algorithm available in the library.

We present a case study of a neural network algorithm based on {\garnet}, applied to a task of identifying the nature of an incoming particle and simultaneously estimating its energy from the energy deposition patterns in a simulated imaging calorimeter. 
%We first describe a benchmark offline implementation of the algorithm, which runs on processors (CPUs and GPUs) without resource and latency constraints, and sets the best achievable accuracy of the model. 
%We then introduce a \textit{compressed} implementation of the algorithm, where parts of the neural network weights are quantized to reduce the necessary compute resource. 
%The two implementations are translated to FPGA firmware using {\hlsfml}, and the inference accuracy of the translation results are compared with the implementations running on processors. 
The inference accuracy of the firmware implementation of the algorithm is compared against its offline counterpart running on processors (CPUs and GPUs).
Latency and resource utilization of the translated FPGA firmware are reported, along with a discussion on their implications for real-world usage of similar algorithms.

This paper is structured as follows. 
In Section~\ref{sec:related}, we briefly recount related work.
Section~\ref{sec:challenges} defines the main problem by outlining the challenges in designing a graph network compatible with L1T latency and resource constraints. 
Section~\ref{sec:garnet} describes how {\garnet} addresses these challenges, and introduces a simplified form of the algorithm with a better affinity to a firmware implementation. 
The case study using a calorimeter simulation is presented in Section~\ref{sec:case_study}, with detailed descriptions of the task setup, model architecture, training results, and the summary of FPGA firmware synthesis. 
Finally, conclusions are given in Section~\ref{sec:conclusion}.

\section{Related work}
\label{sec:related}

Graph neural networks are gaining interest in HEP applications, mainly due to their intrinsic advantage in dealing with sparse input datasets, which are very common in HEP. 
A recent review of applications of GNNs to HEP problems may be found in Ref.~\cite{Shlomi:2020gdn}. 
In particular, dynamic GNNs~\cite{Qasim:2019otl,DGCNN,Kieseler:2020wcq,Gray:2020mcm} are relevant for particle reconstruction tasks, such as tracking~\cite{ExaTrkX} and calorimetry~\cite{Qasim:2019otl}.

Development of ML models deployable to FPGA-based L1T systems is helped by tools for automatic network-to-circuit conversion such as {\hlsfml}.
Using {\hlsfml}, several solutions for HEP-specific tasks (e.g. jet tagging) have been provided~\cite{Duarte:2018ite,Summers:2020xiy,DiGuglielmo:2020eqx,Coelho:2020zfu}, exploiting models with simpler architectures than what is shown here.
This tool has been applied extensively for tasks in the HL-LHC upgrade of the CMS L1T system, including an autoencoder for anomaly detection, and DNNs for muon energy regression and identification, tau lepton identification, and vector boson fusion event classification~\cite{p2l1ttdr}. 
However, prior to this work, GNN models had not yet been supported by {\hlsfml}. 
To the best of our knowledge, the present work is the first demonstration of GNN inference on FPGAs for a HEP application.

Outside of HEP, hardware and firmware acceleration of GNN inference, and graph processing in general, has been an active area of study in recent years, motivated by the intrinsic inefficiencies of CPUs and GPUs when dealing with graph data~\cite{besta2019graph,gui2019survey}. 
Refs.~\cite{10.1145/3373087.3375312, yan2020hygcn,9218751,geng2020awbgcn,kiningham2020grip,6861577,10.1145/3007787.3001155} describe examples of GNN acceleration architectures. 
Refs.~\cite{10.1145/3373087.3375312,yan2020hygcn,9218751,geng2020awbgcn} are specific to the graph convolutional network (GCN)~\cite{kipf2017semisupervised}, while the graph inference processor (GRIP) architecture in Ref.~\cite{kiningham2020grip} is efficient across a wide range of GNN models. 
All five architectures are designed for processing graphs with millions of vertices under a latency constraint (10--1000$\microsec$ or more) that is less stringent than in the HEP L1T environment (less than 1$\microsec$), and are thus not directly applicable to our use case. 
Refs.~\cite{6861577,10.1145/3007787.3001155} present frameworks that automatically generate register-transfer level (RTL) implementations for graph computations according to user-defined configurations. 
While these frameworks are applicable to various graph processing tasks, they require the user to specify the design in highly specific nonstandard format, rather than a standard serialized ML model as in our implementation.

\section{General requirements and challenges}
\label{sec:challenges}

In the framework of Ref.~\cite{battaglia2018relational}, a graph is a triplet $(\mathcal{V}, \mathcal{E}, \mathcal{U})$, where $\mathcal{V}$ is a set of entities (vertices) each possessing some attributes in a fixed format, $\mathcal{E}$ is a set of pairwise relations (edges) between the elements in $\mathcal{V}$, potentially possessing some additional attributes, and $\mathcal{U}$ are global (graph-level) attributes. 
While a GNN can be any neural network that acts on such graphs, in this work we specifically consider graph networks (GN)~\cite{battaglia2018relational}, i.e., architectures that consist of repeatable graph-to-graph mapping blocks (GN blocks). 
Each GN block performs some combination of operations such as edge feature transformation, aggregation of neighbors' features at each vertex, vertex feature transformation, global aggregation of edge and vertex features, and global feature transformation. 
A GN takes a graph as an input sample, where the cardinality of $\mathcal{V}$ may differ sample to sample, and infers its properties, which may be anything from a global scalar, such as a classification label of the sample, to new edge attributes. 
%Repeatable mapping blocks grants a high degree of flexibility in the network design. In particular, the networks may be deep, in combination with the 

To be usable as a part of an LHC L1T system, an algorithm must execute within $\mathcal{O}(1)\microsec$ and have the throughput to accept all inputs from each beam crossing every {25\nanosec}.
Time-multiplexing, whereby $N$ copies of the algorithm accept inputs from $N$ different beam crossings, may be used to decrease the throughput requirement by a factor of $N$.
Additionally, there is a practical constraint that the firmware implementation should fit in the FPGA resources of the system, i.e., utilize the resources such as digital signal processing units (DSPs), look-up tables (LUTs), flip-flips (FFs), and block RAM (BRAM) within the limits of chips available on the market. 
%Regionizing, in which the detector is divided into $N$ regions that are processed by $N$ instances of the algorithm, may be used to relax the resource consumption requirement per instance.
Satisfying these requirements with a GNN can be challenging for multiple reasons listed below.
\begin{itemize}
  \item \textbf{Model depth}: Within each GN block, vertices exchange information with other directly connected vertices or with global attributes. 
  Therefore, to expand the receptive field of each vertex beyond the nearest neighbors, multiple GN blocks must be repeated in the network. 
  Given that various transformations within each GN block are often themselves multilayer perceptrons (MLPs), GNN models tend to be quite deep. 
  Deep networks go against the latency requirement, as each perceptron layer uses at least one clock cycle on an FPGA under a straightforward implementation, and also against the resource usage requirement, because MLPs utilize multiplications heavily.
  \item \textbf{Input size}: Typically, for problems where the application of GNNs is interesting, the cardinality of $\mathcal{V}$ is at least $\mathcal{O}(10^2)$. 
  Even with the high degree of parallelism of FPGAs, due to finiteness of the compute resource, such large input will have to be processed serially to a certain extent, increasing the latency and the interval before a new input can be accepted, known as the initiation interval (II). 
  Longer IIs lead to lower throughput values.
  \item \textbf{Memory usage}: Related to the problem of the input size, if the algorithm requires temporary retention of features for all vertices or edges, memory usage may be prohibitive for an FPGA firmware implementation.
  \item \textbf{Memory access pattern}: Except for certain cases, algorithms that have both $\mathcal{V}$ and $\mathcal{E}$ in the input usually require random memory access, for example when reading or writing features of vertices at the ends of the edges. 
  This poses a challenge in FPGA firmware design not only because it implies that there needs to be a large enough memory bank to store all vertex and/or edge data, but also because random memory access itself is a costly operation~\cite{besta2019graph}. 
  The exceptions include when $\mathcal{E}$ is trivial ($\mathcal{E} = \emptyset$ or when the graph is complete) and when all samples have an identical graph topology. 
  In such cases, the memory access pattern of the algorithm is known at compile time and therefore can be statically scheduled in the FPGA firmware.
\end{itemize}

The case of $\mathcal{E} = \emptyset$ is a rather extreme solution to the last challenge, but it is also attractive in terms of memory usage. 
In fact, even without explicit input edge features, a GNN can infer regional and non-local properties of the graph by globally gathering the vertex features and then scattering the gathered information back to the vertices.
This information flow can also be mediated by a learnable attention mechanism~\cite{velikovi2017graph}. 
The attention mechanism suppresses information from vertices that are considered unimportant, effectively forming ``soft'' edges among the unsuppressed vertices.

In the next section, we study a GNN architecture with these exact properties, then discuss the modifications to the architecture to make it suitable for an FPGA firmware implementation.

\section{A simplified {\garnet} layer in the {\hlsfml} framework}
\label{sec:garnet}

\begin{figure}[t!]
    \centering
    \includegraphics[width=0.6\textwidth]{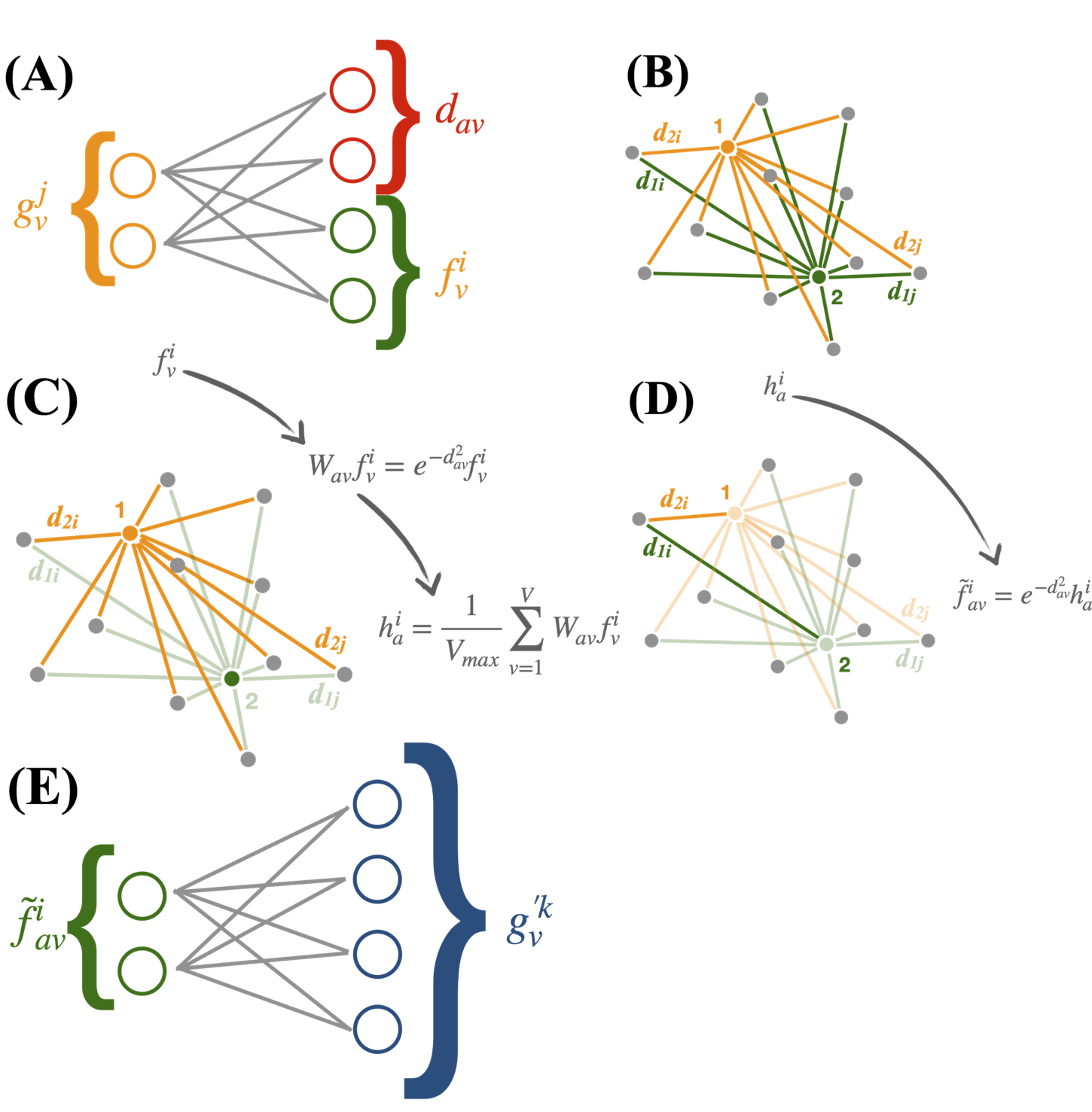}
    \caption{Processing flow of the modified {\garnet} algorithm: (A) The input features ($g^{j}_{v}$) of each vertex are processed by a linear network, that returns a new set of features ($f_v^i$) and its distance from the $S$ aggregators ($d_{av}$). 
    (B) A graph is built in the learned space, using the $d_{av}$ distances. (C) A message is gathered by each aggregator, as a weighted sum across the vertices of $f_v^i$, with $W_{av} = \exp(-d_{av}^2)$ as weights. (D) A message from each aggregator ($\tilde{f}^{i}_{av}$) is passed back to each vertex, with the same $W_{av}$ weight. (E) The aggregated outputs of each vertex are given as input to a neural network, which returns the learned representation.\label{fig:garnet_flow}}
\end{figure}

In this work, we consider \garnet~\cite{Qasim:2019otl} as a specific example of GNN. 
A {\garnet} layer is a GN block that takes as input a set of $V$ vertices, each possessing {\FIN} features, and returns the same set of vertices with {\FOUT} features. 
In a {\garnet} layer, {\FIN} features of each vertex are encoded into an internal representation and gathered at $S$ \textit{aggregators}.
A distance parameter between each of the aggregators and vertices is also computed from the vertex attributes. 
Information gathered at the aggregators are then sent back to individual vertices and decoded into {\FOUT} features. 
Communications between the vertices and aggregators are weighted by a decreasing function of the distance parameter, implementing an attention mechanism that allows the network to learn a dynamic, nontrivial graph structure from the vertex input alone.

The original {\garnet} algorithm, while already using less compute and memory resource than other similar GNN architectures in Ref.~\cite{Qasim:2019otl,DGCNN}, is still challenging to implement as fast and high-throughput FPGA firmware. 
The biggest problem arises from the use of the input feature vector as a part of the input to the decoder, which requires retention of the input data until the last steps of the algorithm. 
An immediate consequence of this requirement is a longer II, because processing of new samples cannot start while the input data for the current sample is still in use. 
Furthermore, the input feature vector is already used to compute the distance parameter as well as the internal representation of each vertex, and therefore a reuse of the input in the decoder creates a complex data flow, restricting the options for pipelining the algorithm.

We therefore designed a modified {\garnet} algorithm with a simplified processing flow:
\begin{itemize}
    \item \textbf{Input transformation} (Figs.~\ref{fig:garnet_flow}(A)~and~(B)): An encoder network converts the features $g^{j}_{v} \thickspace (j=1,\dots,\FIN)$ of the $v^\mathrm{th}$ vertex $(v=1,\dots,V)$ into an internal \textit{learned representation} vector $f^{i}_{v} \thickspace (i=1,\dots,\FLR)$. 
    In parallel, another network (distance calculator) also acts on $g^{j}_{v}$ and computes the distance parameters $d_{av} \thickspace (a=1,\dots,S)$ between the vertices and the $S$ aggregators. 
    Implicitly, this means that a complete bipartite graph with $VS$ edges is built from $\mathcal{V}$ and $\mathcal{S}$, where $\mathcal{S}$ is the set of aggregators (Fig.~\ref{fig:garnet_flow}(B)).
    The encoder and distance calculator networks are both single-layer perceptrons with linear activation functions, so one can write them as linear transformations
    \begin{equation} 
        f^{i}_{v} = \sum_{j=1}^{\FIN} w^{i}_{j} g^{j}_{v} + b^{i} \label{eqn:encoder}
    \end{equation}
    \begin{equation}
        d_{av} = \sum_{j=1}^{\FIN} \alpha_{aj} g^{j}_{v} + \beta_{a}~,
    \end{equation}
    where $(w^{i}_{j}, b^{i})$ and $(\alpha_{aj}, \beta_{a})$ are the kernels and biases of the encoder and distance calculator networks, respectively.
    \item \textbf{Aggregation} (Fig.~\ref{fig:garnet_flow}(C)): The learned representation vectors $f^{i}_{v}$ of the vertices are weighted by a potential function $W_{av} = \exp(-d_{av}^2)$ and averaged across the vertices. 
    In other words, the $i$th averaged feature $h^{i}_{a}$ of aggregator $a$ is written as
    \begin{equation} \label{eqn:aggregation}
        h^{i}_{a} = \frac{1}{\Vmax}\sum_{v=1}^{V} W_{av} f^{i}_{v}.
    \end{equation}
    The factor {\Vmax} in the denominator is the maximum possible value for the vertex multiplicity $V$ (as $V$ may have a different value for each input sample). 
    Through this normalization by a common factor, the information about the size of the sample (cardinality of $\mathcal{V}$) is effectively encoded into $h^{i}_{a}$.
    \item \textbf{Output transformation} (Figs.~\ref{fig:garnet_flow}(D)~and~(E)): The aggregated features are sent back to the vertices using the same weights as
    \begin{equation}
        \tilde{f}^{i}_{av} = W_{av} h^{i}_{a},
    \end{equation}
    and then transformed by a single-layer decoder network with linear activation function into the final output representation $g'^{k}_{v} \thickspace (k=1,\dots,\FOUT)$. 
    With the kernel $u$ and bias $c$ of the decoder, this is written as
    \begin{equation} \label{eqn:decoder}
        g'^{k}_{v} = \sum_{i=1}^{\FLR}\sum_{a=1}^{S} u^{k}_{ia} \tilde{f}^{i}_{av} + c^{k}.
    \end{equation}
\end{itemize}

This simplified algorithm differs from the original design in the following ways. 
First, only the mean over vertices is computed at the aggregators, whereas the maximum is also used in the original design. 
In other words, the aggregators in the original design have
\begin{equation}
    h'^{i}_{a} = \max_{v} W_{av} f^{i}_{v}
\end{equation}
as an additional set of features. 
Secondly, as already noted, the input feature vector is not used as a part of the input to the decoder network. 
In the original {\garnet} design, the decoder is expressed as
\begin{equation}
    g'^{k}_{v} = \sum_{i=1}^{\FLR}\sum_{a=1}^{S} W_{av} \left( u^{k}_{ia} h^{i}_{a} + u'^{k}_{ia} h'^{i}_{a} \right) + \sum_{i=1}^{\FIN} w'^{k}_{i} g^{i}_{v} + c^{k},
\end{equation}
with additional sets of kernel weights $u'$ and $w'$. 
Finally, the original design applies a nonlinear ($\tanh$) activation function to the decoder, while the simplified version uses a linear activation.
In the specific case considered in the next section, these simplifications result in negligible degradation of the network performance. 
In the remainder of this paper, this simplified version of the algorithm is referred to as {\garnet}.

It is worth pointing out that while the {\garnet} layer uses only linear activation functions for all of the internal neural networks, it can still learn nonlinear functions through the nonlinearity of the potential function $W_{av}$. 
On the other hand, having no nonlinear activation functions allows a compact FPGA firmware implementation of the layer, consisting mostly of multiplications and additions. 
The only substantial computation comes with the exponential function, whose values can be pre-computed with sufficient granularity and stored.

An FPGA firmware implementation of the {\garnet} layer using Vivado~\cite{o2014xilinx} HLS is integrated into the {\hlsfml} library. 
The HLS source code is written in C++ and is provided as a template, from which an HLS function for a {\garnet} layer can be instantiated, specifying the configurable parameters such as $S$, \FLR, and \FOUT. In the following, we provide some noteworthy details of the implementation.

In the HLS source code of \garnet, all quantities appearing in the computation are expressed as either integers or fixed-point numbers with fractional precision of at least eight bits.
In particular, the distance parameter $d_{av}$ is represented with three integer bits, eight fractional bits, and one sign bit. 
During the layer computation, $d_{av}$ is reinterpreted as a 12-bit unsigned integer, which is used to retrieve the corresponding pre-computed value of $W_{av}$ from a table with 4,096 entries.

The processing flow in Eqs.~\eqref{eqn:encoder} to \eqref{eqn:decoder} is compactified in the {\hlsfml} implementation by exploiting the linearity of the encoder, average aggregation, and the decoder. 
Equations~\eqref{eqn:encoder}, \eqref{eqn:aggregation}, and \eqref{eqn:decoder} can be combined into
\begin{equation}
    g'^{k}_{v} = \sum_{a=1}^{S} W_{av} \left( \sum_{j=1}^{\FIN} \tilde{w}^{k}_{ja} G^{j}_{a} + \tilde{b}^{k}_{a} L_{a} \right) + c^{k},
\end{equation}
where
\begin{equation}
    \tilde{w}^{k}_{ja} = \sum_{i=1}^{\FLR} u^{k}_{ia} w^{i}_{j}, \quad \tilde{b}^{k}_{a} = \sum_{i=1}^{\FLR} u^{k}_{ia} b^{i}, \quad 
    G^{j}_{a} = \frac{1}{\Vmax} \sum_{v=1}^{V} W_{av} g^{j}_{v}, \thickspace\thickspace \mathrm{and} \thickspace\thickspace L_{a} = \frac{1}{\Vmax} \sum_{v=1}^{V} W_{av}.
\end{equation}
In particular, the kernel and bias tensors of the encoder and decoder are contracted into $\tilde{w}$ and $\tilde{b}$ at logic synthesis time, resulting in fewer steps to arrive at the output from the input.

With this simplification, the input data from each sample are encoded into $W_{av}$, $G^{j}_{a}$, and $L_{a}$. 
Therefore, a new sample can be processed as soon as the three quantities from the previous sample are computed. 
In other words, the II of the overall {\garnet} layer depends on the number of clock cycles needed to compute the three quantities. 
Furthermore, $G^{j}_{a}$ and $L_{a}$ can be derived trivially from $W_{av}$, making the latency of the computation of the latter the critical determinant of the throughput of the algorithm.

The computation of $W_{av}$ is performed independently on each vertex, and is therefore parallelizable across the vertices. 
In a fully parallelized implementation, there would be {\Vmax} logic units (one unit per vertex) operated simultaneously. 
However, with $V$ typically as large as $\mathcal{O}(10^2)$ or greater, this configuration would consume too much of the FPGA resources and would not fit on a single chip. 
Therefore, the {\hlsfml} implementation of {\garnet} allows a partial parallelization of the algorithm controlled by a parameter called the \textit{reuse factor} (\Rreuse). 
For $\Rreuse > 1$, the logic unit to compute $W_{av}$ is cloned $\Vmax / \Rreuse$ times, such that each unit is reused serially up to {\Rreuse} times. 
This serial reuse is fully pipelined with the local II of one clock cycle. 
The latency $T_{W}$ for computing $W_{av}$ for all vertices is therefore given by
\begin{equation} \label{eqn:w_latency}
    T_{W} = T^{0}_{W} + \Rreuse,
\end{equation}
where $T^{0}_{W} \sim 20$ is the number of clock cycles needed to compute $W_{av}$ for one vertex. The value of $T^{0}_{W}$ depends on the numerical precision of the fixed-point numbers in the computation.

Finally, the kernel and bias of the encoder and the kernel of the decoder can be quantized, such that each element takes only values $-1$, $0$, or $1$ (ternary quantization)~\cite{zhu2016trained}. 
In the quantized version of the algorithm, contracted kernel and bias $\tilde{w}$ and $\tilde{b}$ have elements that are $\mathcal{O}(1)$ integers. 
Multiplication of small integers with fixed-point numbers can be performed in FPGAs using LUTs rather than DSPs, which are usually the more scarce resource. 
Multiplications with LUTs also proceed faster than those with DSPs.

\section{Case study: particle identification and energy regression in an imaging calorimeter}
\label{sec:case_study}

As a case study, the {\hlsfml} implementation of {\garnet} is applied to a representative task for the LHC L1T, namely reconstructing electrons and pions in a simulated 3D imaging calorimeter. 
In the following, we first describe the dataset used for the study, then define the task and the architectures of the ML models, and present the inference performance of the models and the resource usage of the synthesized firmware.

\subsection{Dataset}
\label{subsec:data}

The calorimeter is a multi-layered full-absorption detector with a geometry similar to the one described in Ref.~\cite{Qasim:2019otl}. The detector is made entirely of tungsten, which is considered as both an absorber and a sensitive material, and no noise or threshold effects in the readout electronics are simulated.
While this homogeneous calorimeter design is not a faithful representation of a modern sampling calorimeter, this simplification allows us to evaluate the performance of the ML models decoupled from detector effects.

\begin{figure}
    \centering
    \includegraphics[width=0.6\textwidth]{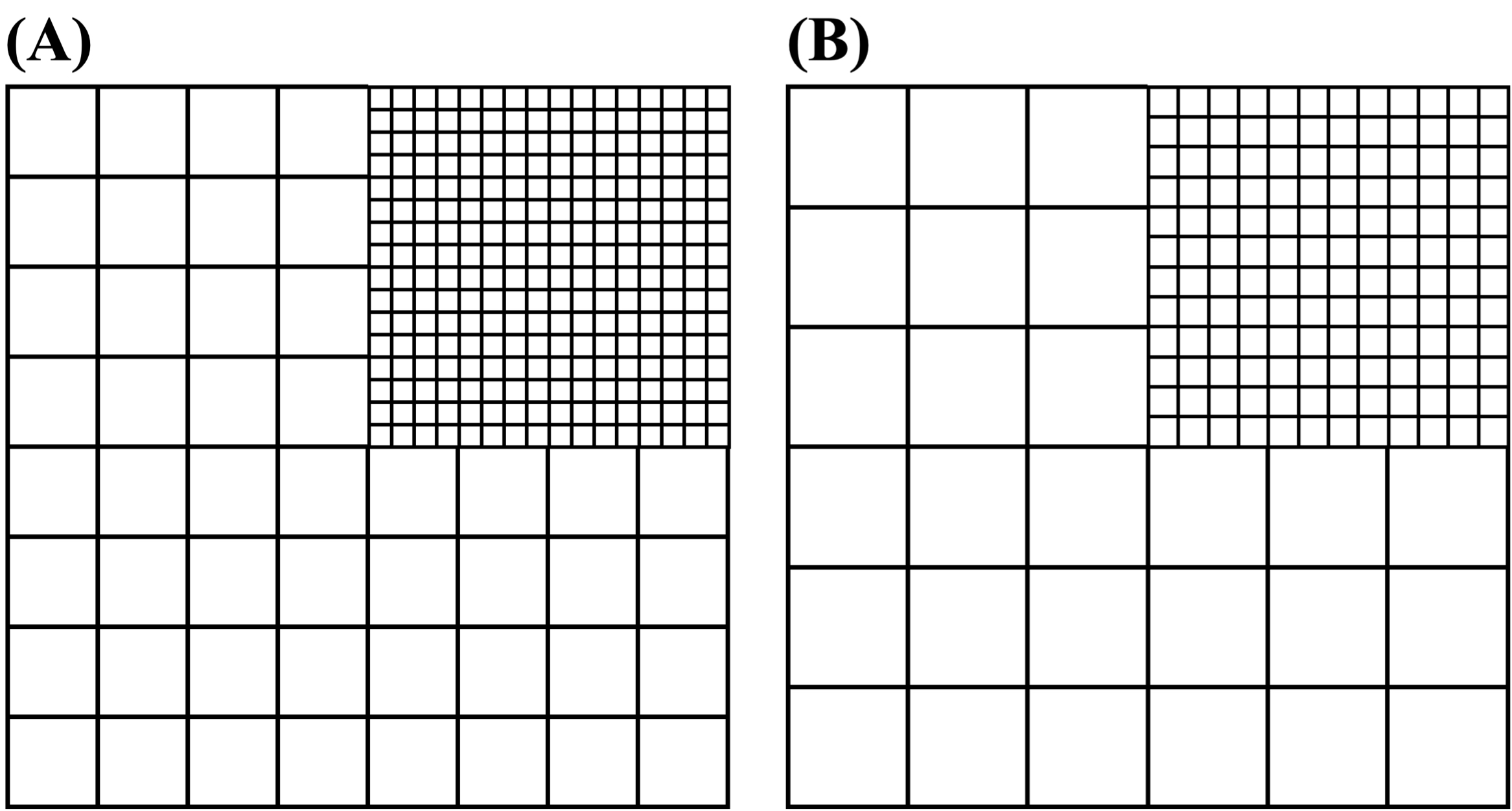}
    \caption{Schematics of the high-granularity and low-granularity regions of the (A) electromagnetic and (B) hadron layers.}
    \label{fig:layer_geometry}
\end{figure}

The calorimeter extends 36~cm in $x$ and $y$ and has a total depth in $z$ of 2~m, corresponding to approximately 20 nuclear interaction lengths and 170 radiation lengths. 
The coordinate origin is placed at the center of the front face of the calorimeter. 
The calorimeter is segmented into 50 layers along $z$, with each layer divided into small square cells in the $x$-$y$ plane, forming a three-dimensional imaging detector. 
Cells are oriented so their sides are parallel to the $x$ and $y$ axes. 
Tiling of the cells in each layer is uniform except for in one quadrant, where the cell sides are half as long as those in the other area. 
The aim of the tiling is to incorporate the irregularity of the geometry of a real-life particle physics calorimeter.
The quadrant with smaller cells and the remainder of the layer are respectively called the high granularity (HG) and low granularity (LG) regions. 
The first 25 layers in $z$ correspond to the electromagnetic calorimeter, with a layer thickness of 1~cm and cell dimensions of 2.25~cm~$\times$~2.25~cm in the HG region (4.5~cm~$\times$~4.5~cm in LG). 
The remaining 25 layers correspond to the hadron calorimeter, with a layer thickness of 7~cm and cell dimensions of 3~cm~$\times$~3~cm in the HG region (6~cm~$\times$~6~cm in LG). 
Schematics of the cell tiling in the electromagnetic and hadron parts are shown in Fig.~\ref{fig:layer_geometry}.
The geometry and the detector response to particles are simulated using \textsc{Geant4}~\cite{Agostinelli:2002hh}. 

Each event used in this study contains a high-energy \textit{primary} particle and low-energy \textit{pileup} particles, which represent backgrounds from simultaneous additional proton-proton interactions.
The primary particle is either an electron ($\mathrm{e}^{-}$) or a charged pion ($\pi^{\pm}$), shot at the calorimeter with momentum aligned along the $z$ axis, i.e., perpendicular to the front face of the calorimeter. 
The $x$ and $y$ coordinates of the particle's origin are randomly sampled according to a uniform distribution in a 10~cm~$\times$~10~cm region centered at $x=y=0$. 
Following this procedure, we aim to mimic a realistic situation in which the actual calorimeter extends to a much larger surface and the area covered by the geometry used in this study represents a portion of it. 
The value of the particle momentum is drawn randomly for each event from a uniform distribution between 10~GeV and 100~GeV. 
The pileup particles consist of photons ($\gamma$) and $\pi^{\pm}$. 
The number of pileup particles is randomly sampled from a Poisson distribution with a mean of 40, with the $\pi^{\pm}$ multiplicity fixed to twice the $\gamma$ multiplicity. 
This setup approximates the flux of pileup particles expected at a pseudorapdity $\eta=2$ in a $\Delta \eta \times \Delta \phi = 0.4 \times 0.4$ patch of the forward region of an LHC detector during the High-Luminosity LHC (HL-LHC) phase~\cite{Apollinari:2284929}. 
The momentum direction and the window of origin of the pileup particles are the same as the primary particle. 
The momentum value of the pileup particles is sampled from a Landau distribution with $\mu = 0.6$~GeV and $c=0.5$~GeV, in a range of 0~to~20~GeV.

The output of the simulation for each event is the array of total energy deposition values by the particles at individual detector cells (hits).
Energy depositions by the particles in the homogeneous calorimeter are recorded exactly, i.e., the detector output does not require calibration and is not affected by stochastic noise.

\begin{figure}[htbp]
    \centering
    \includegraphics[width=0.9\textwidth]{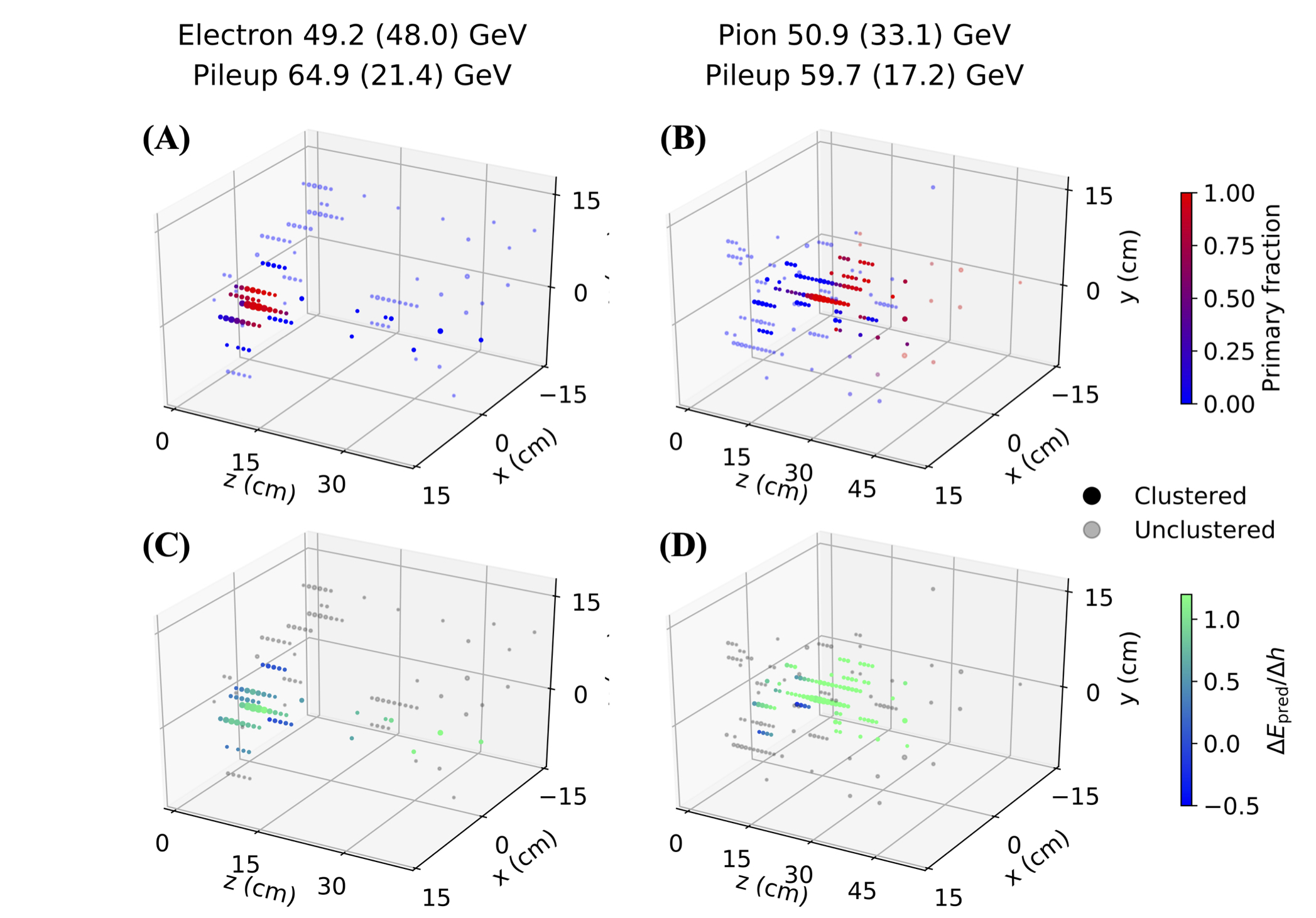}
    \caption{Examples of electron (A, C) and pion (B, D) events. 
    Values in parentheses in the graph titles are the respective energy depositions contained in the cluster around the seed hit. 
    Points represent hits in the detector, with their coordinates at the center of the corresponding detector cells and the size of the markers proportional to the square root of the hit energy. 
    Opaque points are within the cluster, while the translucent ones are not. 
    In (A) and (B), the point color scale from blue to red corresponds to the primary fraction (see Section~\ref{subsec:data} for definition). 
    In (C) and (D), the color scale from blue to green corresponds to $\DEDh$, which is an indication of the importance the neural network model places to individual hits for energy regression. 
    See Section~\ref{subsec:training_result} for details.}
    \label{fig:event_display}
\end{figure}

In an L1T system, hits containing energy depositions from a potentially interesting particle would be identified through a low-latency clustering algorithm. 
The clustering algorithm used in this study mimics the one planned for the L1T system of the HGCAL detector in CMS~\cite{CMS:2293646}. 
In this approach, the hit with the largest energy deposition in the event is elected to be the seed, and the cluster consists of all hits contained in a cylinder whose axis passes through the center of the seed cell and extends along the $z$ direction. 
The radius of the cylinder is set at 6.4~cm so that the resulting cluster contains 95\% of the energy of the primary particle for 50\% of the pion events. 
Because electromagnetic showers have a narrower energy spread than hadronic showers in general, all of the electron events have at least 95\% of the energy contained in the same cylinder. 
Typical events with momenta of the primary particles around 50~GeV and the total pileup energy close to the median of the distribution are shown in Fig.~\ref{fig:event_display}(A) and (B). 
The hits in the figure are colored by the fraction of the hit energy due to the primary particle (primary fraction, \fprim) to help the visualization.

The actual dataset used in this study thus contains one cluster per sample, given as an array of hits in the cluster, and one integer indicating the number of hits in the sample. Only the hits with energy greater than 120~MeV are considered. 
Each cluster contains at most 128 hits, sorted by hit energy in decreasing order. 
Note that sorting of the hit has no effect on the neural network, and is only relevant when truncating the list of hits to consider smaller clusters, as explored later. 
In fact, 0.2\% of the events resulted in clusters with more than 128 hits, for which the lowest energy hits were discarded from the dataset. 
Each hit is represented by four numbers, corresponding to the hit coordinates, given in $x$, $y$, and $z$, and energy. 
The $x$ and $y$ coordinates are relative to the seed cell. 
The dataset consists of 500,000 samples, split evenly and randomly into $\mathrm{e}^{-}$ and $\pi^{\pm}$ events, stored as \textsc{NumPy}~\cite{numpy,numpy2} arrays in \textsc{HDF5} format~\cite{hdf5}. 
The dataset together with the ground truth information is available on the Zenodo platform~\cite{iiyama_yutaro_2020_3888910}.

%To regress the energy of a particle candidate, we limit the study to charged pions. Due to their strongly irregular showers and to their a-priori unknown breakdown in hadronic and electromagnetic energy, these particles represent the most challenging energy-regression use case. A correct determination of $\pi^\pm$ energy profits the most from accurate identifications of the electromagnetic and hadronic shower components on a shower-by-shower basis. Accomplishing this result is particularly challenging in presence of pileup, calling for the development of advanced reconstruction algorithms and specifically tuned detector geometries~\cite{} {\bf C. Neubuser, JK, paper in preparation (only if we are faster than this one here;))}. 

%\begin{table}[ht!]
%    \centering
%    \begin{tabular}{|c|c|c|}
%    \hline
%        Layer &  Cells in x & Cells in y \\ 
%        
%    \hline
%    \end{tabular}
%    \caption{Cell granularity per calorimeter layer.\label{tab:geometry}}
%\end{table}

\subsection{Task and model architecture}
\label{subsec:task_and_architecture}

The task in this study is to identify the nature of the primary particle and to simultaneously predict its energy, given the hits in the cluster. 
The ability to reliably identify the particle type and estimate its energy at the cluster level in a local calorimeter trigger system greatly enhances the efficacy of high-level algorithms, such as particle-flow reconstruction~\cite{Buskulic:1994wz,Sirunyan:2017ulk,Aaboud:2017aca}, downstream in the L1T system.
However, because of the distortion of the energy deposition pattern in the cluster due to pileup, particle identification based on collective properties of the hits, such as the depth of the energy center of mass, can achieve only modest accuracy. 
Furthermore, only half of the pion events have 95\% of the energy deposition from the pion contained in the cluster, requiring substantial extrapolation in the energy prediction. 
This task is thus both practically relevant and sufficiently nontrivial as a test bench of a {\garnet}-based ML model.

The architecture of the model is as follows. First, the input data represented by a two-dimensional array of $\Vmax \times \FIN$ numbers per cluster are processed by a stack of three {\garnet} layers. 
The parameters $(S, \FLR, \FOUT)$ for the first two layers are $(4, 8, 8)$ and for the last layer are $(8, 16, 16)$. 
The output of the third {\garnet} layer is averaged across the vertices for each of the 16 features. 
The resulting array of 16 numbers is then passed through two fully connected layers with 16 and 8 nodes and ReLU~\cite{agarap2018learning} activation. 
Data flow is split into two branches in the final step. 
The first branch consists of a fully connected layer with a single node, whose output is activated by a sigmoid function and is interpreted as the classification prediction, i.e., the predicted probability that the primary particle is an electron. 
The other branch also consists of a single-node fully connected layer, but with a linear activation of the output, which is interpreted as the predicted value of the energy of the particle.

This model is built in \keras~\cite{chollet2015keras}, using the corresponding implementation of {\garnet} available in Ref.~\cite{calographNN}. 
In total, the model has 3,402 trainable parameters (2,976 in the three {\garnet} layers), whose values are optimized through a supervised training process using the Adam optimizer~\cite{kingma2014method}. 
Input is processed in batches of 64 samples during training. 
The overall objective function that is minimized in the training is a weighted sum of objective functions for the classification and regression tasks:
\begin{equation}
   \mathcal{L} = \beta \mathcal{L}_{\mathrm{class}} + (1-\beta) \mathcal{L}_{\mathrm{reg}}
\end{equation}
with $\beta = 0.01$.  The objective function for classification $\mathcal{L}_{\mathrm{class}}$ is the binary cross entropy in each batch between the truth labels (electrons are represented by 1 and pions by 0) and the classification output of the model. 
The objective function for regression $\mathcal{L}_{\mathrm{reg}}$ is the batch mean of the relative squared error
\begin{equation}
    \mathcal{L}_{\mathrm{reg}} = \left[ (\Epred - \Etrue) / \Etrue \right]^2,
\end{equation}
where $\Epred$ and $\Etrue$ are the predicted and true energies of the primary particle, respectively. 
The training is performed on 400,000 training and 100,000 validation samples over a few hundred epochs, with early stopping when the value of the objective function does not improve for ten consecutive epochs. 
Keeping the full training dataset on RAM and using two NVIDIA GeForce RTX 2080 Ti GPUs in parallel, each epoch takes roughly 30 seconds to process.

%Scan range was too narrow (\FLR = 16, \FOUT = 16 in the last layer ended up giving a better performance) so the current model is not optimized
%The model architecture above is determined as a result of a scan of the following hyperparameter values: the number of {\garnet} layers, between 1 and 5; the values of {\FLR} and {\FOUT} for each of the {\garnet} layers, between 4 and 8. corresponds to the configuration with the smallest validation purpose.

Additionally, we prepare a model in which the encoders and decoders of the {\garnet} layers are quantized as ternary networks using \qkeras~\cite{qkeras,Coelho:2020zfu}, which performs quantization-aware training with the straight-through estimator by quantizing the layers during a forward pass but not a backward pass~\cite{binaryconnect,moons2017minimum,zhou2016dorefanet,Coelho:2020zfu}.
In the following, this model is referred to as the \textit{quantized model}, and the original model as the \textit{continuous model}. 
The quantized model is trained with the same objective function and training hyperparameters as the continuous model.

To evaluate the inference performance of the trained models, reference algorithms are defined separately for the classification and regression subtasks. 
The reference algorithm for classification (\textit{cut-based} classification) computes the energy-weighted mean $\bar{z}$ and standard deviation $\sigma_{z}$ of the $z$ coordinates of the hits,
\begin{equation}
    \bar{z} = \frac{\sum_{i=1}^{V} z_{i} h_{i}}{\sum_{i=1}^{V} h_{i}}
    \quad \mathrm{and} \quad
    \sigma_{z} = \sqrt{\frac{\sum_{i=1}^{V} (z_{i} - \bar{z})^2 h_{i}}{\sum_{i=1}^{V} h_{i}}},
\end{equation}
where $i$ is the index of hits in the cluster and $z_{i}$ and $h_{i}$ are the $z$ coordinate and energy of the $i$th hit. 
The cluster is labeled as an electron if $\bar{z} < \bar{z}^{\mathrm{cut}}$ and $\sigma_{z} < \sigma_{z}^{\mathrm{cut}}$, where $\bar{z}^{\mathrm{cut}}$ and $\sigma_{z}^{\mathrm{cut}}$ are predefined thresholds.
Pions, and hadrons in general, tend to penetrate deeper in an absorbing detector and create showers of secondary particles with a larger transverse size than electrons and photons. 
For regression, the reference algorithm (\textit{weight-based} regression) predicts the energy of the primary particle through a formula
\begin{equation}
\label{eqn:regression_ref}
    \Epred^{\mathrm{ref}} = \sum_{i=1}^{V} w_{l(i)} \left( h_{i} + b_{l(i)} \right),
\end{equation}
where $l(i)$ is the detector $z$ layer of hit $i$. Parameters $\{w_{l}, b_{l}\} \thickspace (l=1,\dots,50)$ are determined by minimizing $\mathcal{L}_{\mathrm{reg}}$ over the training dataset using $\Epred^{\mathrm{ref}}$ as the predicted energy. Particle identification based on the energy deposition profile of the cluster and energy estimation based on weighted sum of hit energies are both common strategies in the conventional, non-ML-based event reconstruction approaches.

\subsection{Training result}
\label{subsec:training_result}

\begin{figure}[htbp]
    \centering
    \includegraphics[width=0.9\textwidth]{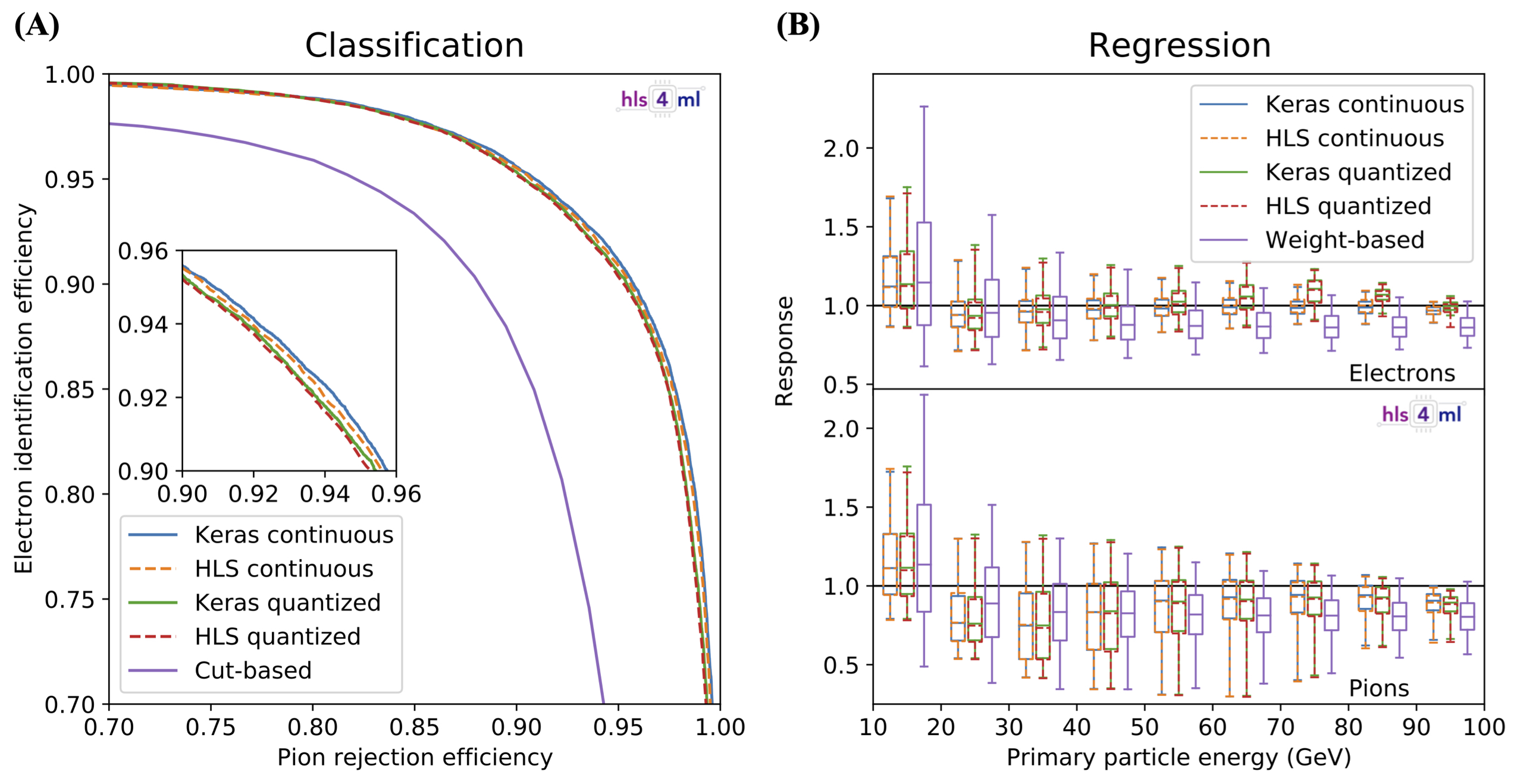}
    \caption{Classification (A) and regression (B) inference performance of the continuous and quantized {\garnet}-based models and the reference algorithms. 
    Results from the {\keras} and HLS implementations are shown for the {\garnet}-based models.
    The classification performance is quantified with a ROC curve of electron identification efficiency versus pion rejection efficiency.
    The inset in (A) shows a close-up view of the efficiency range 0.90--0.96 for both axes.
    The regression performance is quantified as the response ($\Epred / \Etrue$) in 10~GeV bins of $\Etrue$.
    The horizontal line in the box corresponds to the median of the distribution, the top and bottom of the box to the upper and lower quartiles, and the upper and lower ends of the whiskers to the 95th and 5th percentiles.}
    \label{fig:physics_results}
\end{figure}

Performance of the trained continuous and quantized models, evaluated using the validation sample, are shown in Fig.~\ref{fig:physics_results}.
For each ML model, the inference results based on the original {\keras} model and the HLS model, converted using {\hlsfml}, are shown.
The HLS model provides a realistic emulation of the synthesized FPGA firmware.

The classification performance is given in terms of receiver operating characteristic (ROC) curves that trace the electron identification efficiency (true positive fraction) and pion rejection efficiency (true negative fraction) for different thresholds of the classifiers. 
The two {\garnet}-based models perform similarly and better than the cut-based reference in terms of the electron identification efficiency for a given pion rejection efficiency.
A detailed comparison of the four sets of results from the {\garnet}-based models in the inset reveals that the continuous model performs slightly better than the quantized model, and that the difference between the {\keras} and HLS implementations is smaller for the quantized model.

The regression performance is given in terms of the response ($\Epred / \Etrue$). 
Distributions of the response are summarized in 10~GeV bins of $\Etrue$, separately for the continuous model, quantized model, and the weight-based reference. 
In each summary, the horizontal line in the box corresponds to the median of the distribution, the top and bottom of the box to the upper and lower quartiles, and the upper and lower ends of the whiskers to the 95th and 5th percentiles. 
The {\garnet}-based models exhibit narrower spreads of the response distributions in most of the bins, with the continuous model again performing slightly better than the quantized model.

The differences between the {\keras} and HLS implementations are due to the numerical precision in the computation. 
While the former represents all fractional numbers in 32-bit floating-point numbers, the latter employs fixed-point numbers with bit widths of at most 18. 
Consequently, for the quantized model, where the encoder and decoder of the {\garnet} layers employ integer weights for inference, the difference between the two implementations are smaller.

For both subtasks, the {\garnet}-based models generally outperform the reference algorithms. 
The reference algorithm has narrower spread of the response in some energy bins for the regression subtask. 
However, it is important to note that the weights and biases appearing in Eq.~\eqref{eqn:regression_ref} are optimized for a specific pileup profile, while in a real particle collider environment, pileup flux changes dynamically even on the timescale of a few hours. 
In contrast, algorithms based on inference of properties of individual hits, such as the {\garnet}-based models presented in this study, are expected to be able to identify hits due to pileup even under different pileup environments and thus to have a stable inference performance with respect to change in pileup flux. 
Since a detailed evaluation of application-specific performance of {\garnet} is not within the scope of this work, we leave this and other possible improvements to the model architecture and training to future studies.

To verify that {\garnet} can infer relations between individual vertices without edges $\mathcal{E}$ in the input, 
%as claimed in Section~\ref{sec:challenges},
the following test is performed. 
Using the two events shown in Fig.~\ref{fig:event_display}, the energy of each hit in the clusters is increased one at a time by 10\%, and the inference with the continuous model is performed for each perturbed event. 
If the model has learned to perfectly distinguish the primary particle from pileup at the vertex level, a small change in the energy of a hit from pileup should result in no change in the predicted particle energy. 
In Fig.~\ref{fig:event_display}(C) and (D), each hit in the cluster is colored by the ratio of the change of predicted particle energy and the amount of perturbation ($\DEDh$). 
While some hits with $\fprim = 0$ appear with $\DEDh > 0$, a general correspondence between {\fprim} and {\DEDh} is observed. 
The occurrence of $\DEDh > 1$ is expected, given the extrapolation required to predict the full particle energy from the energy of the hits included in the cluster. 
With this test, we are able to probe how the {\garnet}-based model is learning the structure of the graph.
%associating the vertices with each other.

\subsection{Model synthesis and performance}
\label{subsec:model_synthesis}

The latency, II, and resource usage of the FPGA firmware synthesized from the HLS implementations are summarized in Table.~\ref{tab:synthesis_summary}.
Vitis Core Development Kit 2019.2~\cite{vitis} is used for synthesis, with a Xilinx Kintex UltraScale FPGA (part number \texttt{xcku115-flvb2104-2-i}) as the target device and a clock frequency of 200~MHz.
The reported resource usage numbers reflect the synthesis estimates from Vivado HLS.
The latency and II reported here are the maximum values for samples with full {\Vmax} vertices; the actual HLS implementation allows early termination of the serial reuse of the vertex-processing logic unit for samples with fewer vertices.
The area under the ROC curve (AUC) and overall response root mean square (RMS) are used to summarize the performance.

\begin{table}[hbtp]
    \centering
    \caption{Summary of the latency, II, FPGA resource usage metrics, and inference accuracy metrics of the synthesized firmware. 
    The reported resource usage numbers reflect the synthesis estimates from Vivado HLS.
    The target FPGA is a Xilinx Kintex UltraScale FPGA (part number \texttt{xcku115-flvb2104-2-i}), which has 5,520 DSPs, 663,360 LUTs, 1,326,720 FFs, and 77.8~Mb of BRAM~\cite{datasheet}. 
    The utilized percentage of the targeted FPGA resources are denoted in the square brackets.}
    \label{tab:synthesis_summary}
    \resizebox{\textwidth}{!}{
    \begin{tabular}{ccc|cccccc|cc}
    \multirow{2}{*}{Model} & \multirow{2}{*}{\Vmax} & \multirow{2}{*}{\Rreuse} & Latency & Interval & \multirow{2}{*}{DSP ($10^3$)} & \multirow{2}{*}{LUT ($10^3$)} & \multirow{2}{*}{FF ($10^3$)} & \multirow{2}{*}{BRAM (Mb)} & ROC & Response \\
    & & & (cycles) & (cycles) & & & & & AUC & RMS \\
    \hline
    Continuous & 128 & 32 & 155 & 55 & 3.1 [56\%] & 57 [9\%] & 39 [2.9\%] & 1.8 [2.3\%] & 0.98 & 0.23 \\
    Quantized & 128 & 32 & 148 & 50 & 1.6 [29\%] & 70 [11\%] & 41 [3.1\%] & 1.9 [2.4\%] & 0.98 & 0.24 \\
    \hline
    Quantized & 64 & 16 & 99 & 34 & 1.6 [29\%] & 63 [9\%] & 38 [2.9\%] & 1.8 [2.3\%] & 0.96 & 0.24 \\
    Quantized & 32 & 8 & 75 & 26 & 1.4 [25\%] & 52 [8\%] & 33 [2.5\%] & 1.8 [2.3\%] & 0.86 & 0.37 \\
    Quantized & 16 & 4 & 63 & 22 & 1.5 [27\%] & 57 [9\%] & 37 [2.8\%] & 1.8 [2.3\%] & 0.64 & 0.36 \\
    \end{tabular}}
\end{table}

Comparing the continuous and quantized models with $\Vmax = 128$, the former has a longer latency and II and consumes substantially more DSPs. 
On the other hand, the quantized model uses more LUTs, mainly for the multiplications in the {\garnet} encoders and decoders, as discussed in Section~\ref{sec:garnet}. 
However, it is known that the expected LUT usage tend to be overestimated in Vivado HLS, while the expected DSP usage tends to be accurate~\cite{DiGuglielmo:2020eqx,Duarte:2018ite}.
The DSP usage of $3.1\times 10^3$ for the continuous model is well within the limit of the target device, but is more than what is available on a single die slice ($2.8\times 10^3$)~\cite{datasheet}. 
The quantized model fits in one slice in all metrics. 
Given the small difference in the inference performance between the two models, it is clear that the quantized model is advantageous for this specific case study.

The latency of the synthesized quantized model at 148 clock periods, corresponding to 740\nanosec, satisfies the LHC L1T requirement of $\mathcal{O}(1)\microsec$ execution. 
However, the II of 50 clock periods (250\nanosec) implies that the logic must be time-multiplexed tenfold to be able to process a single cluster per LHC beam crossing period of 25\nanosec. 
With $\mathcal{O}(100)$ or more clusters expected per beam crossing in the collision environment of HL-LHC, the throughput of the synthesized firmware is therefore inadequate for a reasonably sized L1T calorimeter system with $\mathcal{O}(100)$ FPGAs, and requires down-scoping or implementation improvements.

The simplest down-scoping measure is to reduce the size of the input. 
This is effective because the most prominent factor driving both the latency and the II of the firmware is {\Rreuse} (see Eq.~\eqref{eqn:w_latency}), which in turn is determined by {\Vmax} to be able to fit the logic in a single chip. 
To test how short the II can be made while retaining a reasonable inference performance, additional models with $\Vmax = 64$, 32, and 16 are trained and synthesized into FPGA firmware. 
Clusters with more hits than {\Vmax} are truncated by discarding the lowest energy hits. 
The fraction of truncated clusters for the three {\Vmax} values are 27\%, 85\%, and 99\%, respectively.

The results of synthesis of the additional models are given in the last three rows of Table~\ref{tab:synthesis_summary}. 
The values of FPGA resource usage metrics are similar in all quantized models because the ratio $\Vmax / \Rreuse$ is kept at 4. 
Only a modest degradation of performance is observed by truncating the clusters to $\Vmax = 64$, while the II is reduced by 16 clocks as a direct result of the reduction of {\Rreuse} by the same amount. 
This working point might thus represent a reasonable compromise between the inference performance and throughput. 
Further cluster truncation results in considerable loss of inference accuracy. 
It is also clear that reduction of {\Rreuse} has a diminishing return in terms of shorter II, and improvements to other parts of the algorithm are necessary to further reduce the II.

\section{Conclusion} 
\label{sec:conclusion}

In this paper, we presented an implementation of a graph neural network algorithm as FPGA firmware with $\mathcal{O}(1)\microsec$ execution time. 
General considerations and challenges in implementing graph neural networks for real-time trigger systems at particle collider experiments are outlined, along with how algorithms such as {\garnet} address these issues. 
We then described the simplified version of \garnet, which is now available as a general-purpose graph network layer in the {\hlsfml} library. 
An example use case of a machine learning model based on the simplified version of {\garnet}, applied to data from a simulation of a small imaging calorimeter, is presented. 
The model is able to learn to predict the identity and the energy of the particles detected at the calorimeter with high accuracy, while its firmware implementation executes in {740\nanosec} and fits easily in a commercially available FPGA. 
Although the throughput of the firmware is not sufficient to make the model readily deployable in a submicrosecond, real-time collider trigger system, its variants with reduced input size are shown to have higher throughput with reasonable inference performance. 
These results demonstrate that fast inference of graph neural networks in FPGAs is possible, and with {\hlsfml}, various graph-based machine learning architectures can be automatically translated into firmware.

\section*{Data availability statement}

Simulation data set and the \keras source code used for the case study are available on the Zenodo platform~\cite{iiyama_yutaro_2020_3888910, iiyama_yutaro_2020_3992780}. 
The repository for the source code also includes input files for \hlsfml~\cite{hls4ml} to generate the HLS models described in this paper.

\section*{Author contributions}

All authors listed have made a substantial, direct, and intellectual contribution to the work and approved it for publication.

\section*{Funding}

M.~P., A.~G., K.~W., S.~S., V.~L. and J.~N. are supported by the European Research Council (ERC) under the European Union's Horizon 2020 research and innovation program (Grant Agreement No. 772369).

S.~J., M.~L., K.~P., and N.~T. are supported by Fermi Research Alliance, LLC under Contract No. DE-AC02-07CH11359 with the U.S. Department of Energy (DOE), Office of Science, Office of High Energy Physics.
P.~H. is supported by a Massachusetts Institute of Technology University grant. 
Z.~W. is supported by the National Science Foundation under Grants No. 1606321 and 115164.
J.~D. is supported by DOE Office of Science, Office of High Energy Physics Early Career Research program under Award No. DE-SC0021187.

\section*{Conflict of interest}

The authors declare that the research was conducted in the absence of any commercial or financial relationships that could be construed as a potential conflict of interest.

\section*{Acknowledgments}

We acknowledge the Fast Machine Learning collective as an open community of multi-domain experts and collaborators. 
This community was important for the development of this project. 

\bibliographystyle{lucas_unsrt} %needed for correct ordering
\bibliography{references}

\end{document}